\documentclass[prl,10pt,twocolumn,superscriptaddress,aps]{revtex4-1}

\usepackage{amsmath}
\usepackage{verbatim}
\usepackage{graphicx}

\begin{document}

\title{Study of $\Xi^*$ Photoproduction \\ from Threshold to $W = 3.3$~GeV}


\newcommand*{\OHIOU}{Ohio University, Athens, Ohio  45701}
\newcommand*{\OHIOUindex}{1}
\affiliation{\OHIOU}
\newcommand*{\Juelich}{Institute fur Kernphysik (Juelich), Juelich, Germany}
\newcommand*{\Juelichindex}{2}
\affiliation{\Juelich}
\newcommand*{\CSUDH}{California State University, Dominguez Hills, Carson, CA 90747}
\newcommand*{\CSUDHindex}{3}
\affiliation{\CSUDH}
\newcommand*{\CNU}{Christopher Newport University, Newport News, Virginia 23606}
\newcommand*{\CNUindex}{4}
\affiliation{\CNU}
\newcommand*{\ANL}{Argonne National Laboratory, Argonne, Illinois 60439}
\newcommand*{\ANLindex}{5}
\affiliation{\ANL}
\newcommand*{\ASU}{Arizona State University, Tempe, Arizona 85287-1504}
\newcommand*{\ASUindex}{6}
\affiliation{\ASU}
\newcommand*{\CC}{Canisius College, Buffalo, New York 14208-1517}
\newcommand*{\CCindex}{7}
\affiliation{\CC}
\newcommand*{\CMU}{Carnegie Mellon University, Pittsburgh, Pennsylvania 15213}
\newcommand*{\CMUindex}{8}
\affiliation{\CMU}
\newcommand*{\CUA}{Catholic University of America, Washington, D.C. 20064}
\newcommand*{\CUAindex}{9}
\affiliation{\CUA}
\newcommand*{\SACLAY}{IRFU, CEA, Universit'e Paris-Saclay, F-91191 Gif-sur-Yvette, France}
\newcommand*{\SACLAYindex}{10}
\affiliation{\SACLAY}
\newcommand*{\UCONN}{University of Connecticut, Storrs, Connecticut 06269}
\newcommand*{\UCONNindex}{11}
\affiliation{\UCONN}
\newcommand*{\DUKE}{Duke University, Durham, North Carolina 27708-0305}
\newcommand*{\DUKEindex}{12}
\affiliation{\DUKE}
\newcommand*{\FU}{Fairfield University, Fairfield CT 06824}
\newcommand*{\FUindex}{13}
\affiliation{\FU}
\newcommand*{\FERRARAU}{Universita' di Ferrara , 44121 Ferrara, Italy}
\newcommand*{\FERRARAUindex}{14}
\affiliation{\FERRARAU}
\newcommand*{\FIU}{Florida International University, Miami, Florida 33199}
\newcommand*{\FIUindex}{15}
\affiliation{\FIU}
\newcommand*{\FSU}{Florida State University, Tallahassee, Florida 32306}
\newcommand*{\FSUindex}{16}
\affiliation{\FSU}
\newcommand*{\Genova}{Universit$\grave{a}$ di Genova, 16146 Genova, Italy}
\newcommand*{\Genovaindex}{17}
\affiliation{\Genova}
\newcommand*{\GWUI}{The George Washington University, Washington, DC 20052}
\newcommand*{\GWUIindex}{18}
\affiliation{\GWUI}
\newcommand*{\ISU}{Idaho State University, Pocatello, Idaho 83209}
\newcommand*{\ISUindex}{19}
\affiliation{\ISU}
\newcommand*{\INFNFE}{INFN, Sezione di Ferrara, 44100 Ferrara, Italy}
\newcommand*{\INFNFEindex}{20}
\affiliation{\INFNFE}
\newcommand*{\INFNFR}{INFN, Laboratori Nazionali di Frascati, 00044 Frascati, Italy}
\newcommand*{\INFNFRindex}{21}
\affiliation{\INFNFR}
\newcommand*{\INFNGE}{INFN, Sezione di Genova, 16146 Genova, Italy}
\newcommand*{\INFNGEindex}{22}
\affiliation{\INFNGE}
\newcommand*{\INFNRO}{INFN, Sezione di Roma Tor Vergata, 00133 Rome, Italy}
\newcommand*{\INFNROindex}{23}
\affiliation{\INFNRO}
\newcommand*{\INFNTO}{INFN, Sezione di Torino, 10125 Torino, Italy}
\newcommand*{\INFNTOindex}{24}
\affiliation{\INFNTO}
\newcommand*{\ORSAY}{Institut de Physique Nucl\'eaire, CNRS/IN2P3 and Universit\'e Paris Sud, Orsay, France}
\newcommand*{\ORSAYindex}{25}
\affiliation{\ORSAY}
\newcommand*{\ITEP}{Institute of Theoretical and Experimental Physics, Moscow, 117259, Russia}
\newcommand*{\ITEPindex}{26}
\affiliation{\ITEP}
\newcommand*{\JMU}{James Madison University, Harrisonburg, Virginia 22807}
\newcommand*{\JMUindex}{27}
\affiliation{\JMU}
\newcommand*{\KNU}{Kyungpook National University, Daegu 41566, Republic of Korea}
\newcommand*{\KNUindex}{28}
\affiliation{\KNU}
\newcommand*{\MISS}{Mississippi State University, Mississippi State, MS 39762-5167}
\newcommand*{\MISSindex}{29}
\affiliation{\MISS}
\newcommand*{\UNH}{University of New Hampshire, Durham, New Hampshire 03824-3568}
\newcommand*{\UNHindex}{30}
\affiliation{\UNH}
\newcommand*{\NSU}{Norfolk State University, Norfolk, Virginia 23504}
\newcommand*{\NSUindex}{31}
\affiliation{\NSU}
\newcommand*{\ODU}{Old Dominion University, Norfolk, Virginia 23529}
\newcommand*{\ODUindex}{32}
\affiliation{\ODU}
\newcommand*{\ROMAII}{Universita' di Roma Tor Vergata, 00133 Rome Italy}
\newcommand*{\ROMAIIindex}{33}
\affiliation{\ROMAII}
\newcommand*{\MSU}{Skobeltsyn Institute of Nuclear Physics, Lomonosov Moscow State University, 119234 Moscow, Russia}
\newcommand*{\MSUindex}{34}
\affiliation{\MSU}
\newcommand*{\SCAROLINA}{University of South Carolina, Columbia, South Carolina 29208}
\newcommand*{\SCAROLINAindex}{35}
\affiliation{\SCAROLINA}
\newcommand*{\TEMPLE}{Temple University,  Philadelphia, PA 19122 }
\newcommand*{\TEMPLEindex}{36}
\affiliation{\TEMPLE}
\newcommand*{\JLAB}{Thomas Jefferson National Accelerator Facility, Newport News, Virginia 23606}
\newcommand*{\JLABindex}{37}
\affiliation{\JLAB}
\newcommand*{\UTFSM}{Universidad T\'{e}cnica Federico Santa Mar\'{i}a, Casilla 110-V Valpara\'{i}so, Chile}
\newcommand*{\UTFSMindex}{38}
\affiliation{\UTFSM}
\newcommand*{\EDINBURGH}{Edinburgh University, Edinburgh EH9 3JZ, United Kingdom}
\newcommand*{\EDINBURGHindex}{39}
\affiliation{\EDINBURGH}
\newcommand*{\GLASGOW}{University of Glasgow, Glasgow G12 8QQ, United Kingdom}
\newcommand*{\GLASGOWindex}{40}
\affiliation{\GLASGOW}
\newcommand*{\VIRGINIA}{University of Virginia, Charlottesville, Virginia 22901}
\newcommand*{\VIRGINIAindex}{41}
\affiliation{\VIRGINIA}
\newcommand*{\WM}{College of William and Mary, Williamsburg, Virginia 23187-8795}
\newcommand*{\WMindex}{42}
\affiliation{\WM}
\newcommand*{\YEREVAN}{Yerevan Physics Institute, 375036 Yerevan, Armenia}
\newcommand*{\YEREVANindex}{43}
\affiliation{\YEREVAN}
 

\author{J.T.~Goetz}  
\affiliation{\OHIOU}
\author{K.~Hicks} 
\affiliation{\OHIOU}
\author {M.C.~Kunkel} 
\affiliation{\Juelich}
\author {J.W.~Price} 
\affiliation{\CSUDH}
\author{D.P.~Weygand}
\affiliation{\CNU}
\author {S. Adhikari} 
\affiliation{\FIU}
\author {S. ~Anefalos~Pereira} 
\affiliation{\INFNFR}
\author {M.~Battaglieri} 
\affiliation{\INFNGE}
\author {I.~Bedlinskiy} 
\affiliation{\ITEP}
\author {A.S.~Biselli} 
\affiliation{\FU}
% ...

\collaboration{The CLAS Collaboration}
\noaffiliation

\date{\today}

\begin{abstract}
The doubly strange $\Xi$ baryons provide an effective way to study a 
puzzle called the missing-baryons problem, where both quark models 
and lattice gauge theory predict more baryon excited states than are 
seen experimentally. 
However, few of these excited states have been observed with any certainty. 
Here, high-mass $\Xi^*$ states have been searched for in photoproduction 
with the CLAS detector, and upper limits for the total cross sections 
have been established from threshold to $W=3.3$~GeV. 
In addition, the total cross sections of the ground state $\Xi^-$(1320) 
and first excited state $\Xi^-$(1530) are presented, extending significantly 
the center-of-mass energy range of previous data.
\end{abstract}

\maketitle


Cascade baryons, also called $\Xi$ states, hold an important place in the 
development of the quark model, 
and continue to be useful to the field of baryon spectroscopy 
\cite{edwards2013flavor}. 
Made from two strange quarks and one light (up or down) quark, the cascade 
baryons come in only two charge states, $\Xi^-$ and $\Xi^0$.  
The $\Xi$ ground state, with $J=1/2$, completes the octet of ground-state 
baryons. 
The first excited state, with $J=3/2$, is part of the baryon decuplet, which 
famously led Gell-Mann to predict the mass of the $\Omega^-$ 
\cite{gell1964eightfold}.

Quark models, both relativistic and non-relativistic \cite{capstick2000quark}, 
along with the chiral-symmetric \cite{glozman1996spectrum} and algebraic 
\cite{bijker2000algebraic} models, all predict many more baryon states than 
have been observed to date. 
This so-called ``missing-baryons problem'' has persisted in light of recent 
lattice QCD calculations \cite{edwards2013flavor}, making the 
measurement of the baryon spectrum a high priority for the understanding of 
QCD theory. 
Looking for $\Xi^*$ states experimentally and taking advantage of the 
known $N^*$ correspondence is the primary motivation for the present study. 
If the number of $\Xi^*$ states found experimentally is also small, it begs 
the question why.

Today, there are better calculations than the quark model for the spectrum 
of excited states of the Cascade baryons, which are done directly using the 
theory of quantum chromodynamics (QCD). 
While these lattice calculations \cite{edwards2013flavor} are still using a 
light quark mass greater than the physical mass, the mass spectrum of 
Cascade baryons can still be extracted and normalized to the $\Omega^-$ 
ground state. 
The resulting pattern is remarkably similar to that for quark-model 
calculations, where states with higher spin have systematically higher mass.  
From lattice methods, seven $\Xi^*$ states have been identified in the 
first resonance region with negative parity corresponding to $L=1$, as 
well as over a dozen excited states at higher mass in the second resonance 
region with positive parity.

Experimentally, it is appealing to look for the excited cascade states 
because they are expected to have a narrow width \cite{riska2003}.  
For example, the ground state $\Delta$ resonance has a width of about 
120~MeV, whereas the $\Xi$(1530)  and the $\Xi$(1690) have widths of about 
10~MeV \cite{pdg}, which are more easily seen above background. 
Furthermore, there should be one $\Xi^*$ for each $N^*$ state, and while 
the $N^*$ states are broad and overlapping, the $\Xi^*$ states are expected 
to be narrow and easily isolated as a peak in the experimental mass spectrum. 
However, the cross section for producing cascade baryons is small, 
especially for photoproduction \cite{price2005exclusive,guo2007cascade}, 
but this situation is well suited to today's high-rate photon beams and 
large-acceptance spectrometers.

A new experiment with the CLAS detector with sufficient photon energy 
and flux to carry out a statistically significant search for $\Xi^*$ 
states above the $\Xi$(1530) was carried out. 
In addition, this is the first time cross sections for the $\Xi$(1320) and 
$\Xi$(1530) have been measured in photoproduction at photon energy 
$E_\gamma > 4$~GeV with good statistics, where the total cross section is 
predicted to level off \cite{nakayama2006photoproduction}. 
We report here total cross sections from fits to angular distributions from 
threshold up to $E_\gamma = 5.4$ GeV.

\begin{figure}[htpb]
    \begin{center}
            \includegraphics[width=\columnwidth]{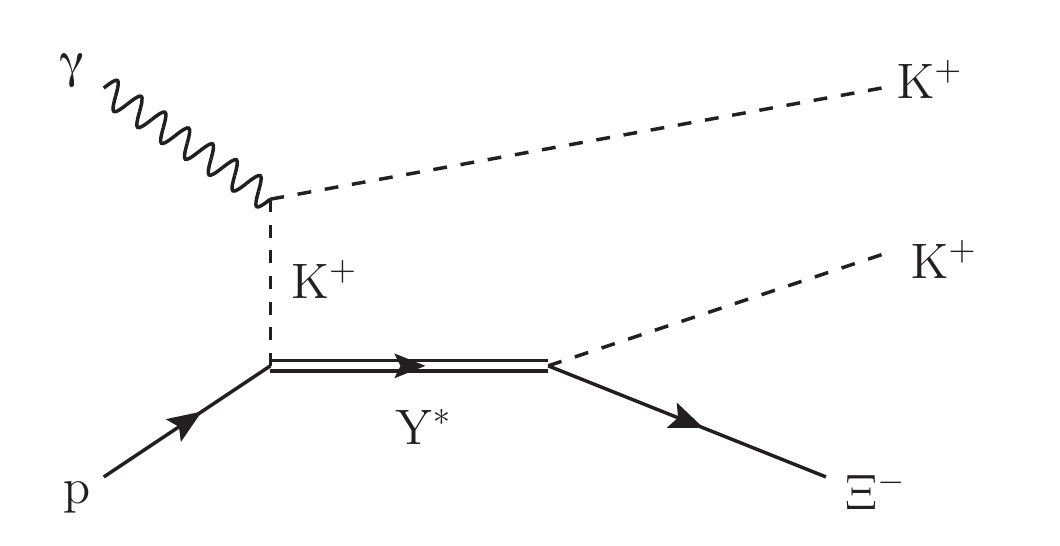}
            \includegraphics[width=\columnwidth]{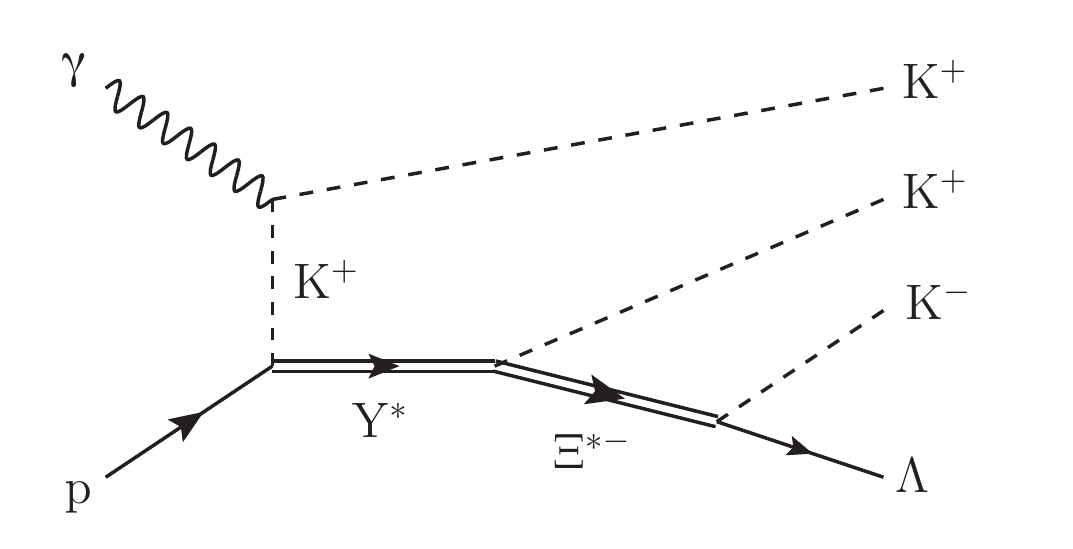}
        \caption{\label{fig:xiprod}Diagrams used by Ref. 
\cite{nakayama2006photoproduction} for photoproduction of the $\Xi^-$ 
ground state (left) and excited states (right) through decay of an 
intermediate hyperon resonance, $Y^*$.}
    \end{center}
\end{figure}

Theoretical calculations for cascade photoproduction have been carried out by 
Nakayama, Oh, and Haberzettl \cite{nakayama2006photoproduction}. 
The production mechanism they propose, shown in Fig.~\ref{fig:xiprod} 
for both ground-state and excited cascades, is a two-step process. 
A high-mass hyperon is made via \mbox{$\gamma p \to K^+ Y^*$} followed by 
a decay branch of the $Y^*$ to \mbox{$K^+ \Xi$}.
Direct production of the $\Xi$ seems unlikely because two 
$s\bar{s}$ quark pairs would need to be created at the 
production vertex (a violation of the OZI rule \cite{pdg}).  
The hadronic coupling constants in this two-step process are unknown, 
and so the theoretical calculations have been normalized to data 
from previous CLAS experiments \cite{guo2007cascade}. 
No calculations have been published for photoproduction of excited 
$\Xi^*$ states.

The $\Xi(1690)$ was seen by the WA89 Collaboration 
\cite{adamovich1998first} with high statistics and more recently by the 
Belle Collaboration \cite{Aubert:2006ux} in $\Lambda_c^+$ decays. 
The $\Xi(1820)$ was seen decaying to $\Lambda\overline{K}^0$ and 
$\Sigma\overline{K}^0$ \cite{biagi1981production} and also from decays to 
$\Lambda K^-$ with $8\sigma$ significance \cite{biagi1987xi_a,biagi1987xi_b}.
Both states have widths of about 25~MeV \cite{pdg} (or even less for 
the $\Xi(1690)$) and should be seen in the present study if the 
photoproduction cross section is sufficiently high.


Electrons from the Continuous Electron Beam Accelerator Facility (CEBAF) 
at Jefferson Lab with beam energy 5.715~GeV were directed onto a thin gold 
radiator foil to produce bremsstrahlung photons. These were collimated onto 
a 40 cm long liquid hydrogen target. 
The CLAS detector \cite{clas} was used for this experiment, known as 
{\rm g12}, which ran in the second quarter of 2008. 
The target center was 90~cm upstream from the center of CLAS to provide 
better acceptance for particles produced at small angles. 
To allow for high luminosity, with a beam current of 60-65~nA, 
a 24-segment scintillator start-counter \cite{clas.st} (ST) 
around the target was used to form a coincidence trigger with the 
time-of-flight \cite{clas.tof} scintillators (TOF) that surrounded 
the outside of CLAS. 
Two ST/TOF pairs of hits in separate sectors of CLAS in 
coincidence with a scattered electron in the bremsstrahlung tagger 
\cite{clas.tagger} were required to satisfy the trigger. 
These conditions, along with several ancillary trigger conditions, 
resulted in a livetime of the data acquisition system of $\sim$87\%. 
A trigger coincidence window of approximately 100~ns resulted in 
about 20-30 recorded photons per event.

In this analysis, events were defined as two $K^+$ particles detected 
in CLAS within 1.0~ns of the photon's vertex time and the two-particle 
vertex within the target volume. 
The $K^+ K^+$ vertex time was calculated using the time at the 
TOF along with its momentum and path length measured in the 
drift chambers. 
In addition, each track was required to have a valid hit in the ST 
within a $\pm 1.6$~ns time window.  
These timing cuts were rigorously calibrated and studied for their 
overall efficiency. 
With these cuts, clean identification of the two charged kaons
became possible.

One additional event selection criterion was applied to the data. 
The mass of each charged particle can be independently calculated from the 
momentum along with the velocity from the TOF. 
At high momenta, the above timing cuts alone become less effective 
at separating kaons from pions.  An additional cut on the calculated mass, 
within 20~MeV of the known kaon mass, was applied to further reduce 
background from misidentified pions.

\begin{figure}[htpb]\begin{center}
\includegraphics[width=1.0\columnwidth]{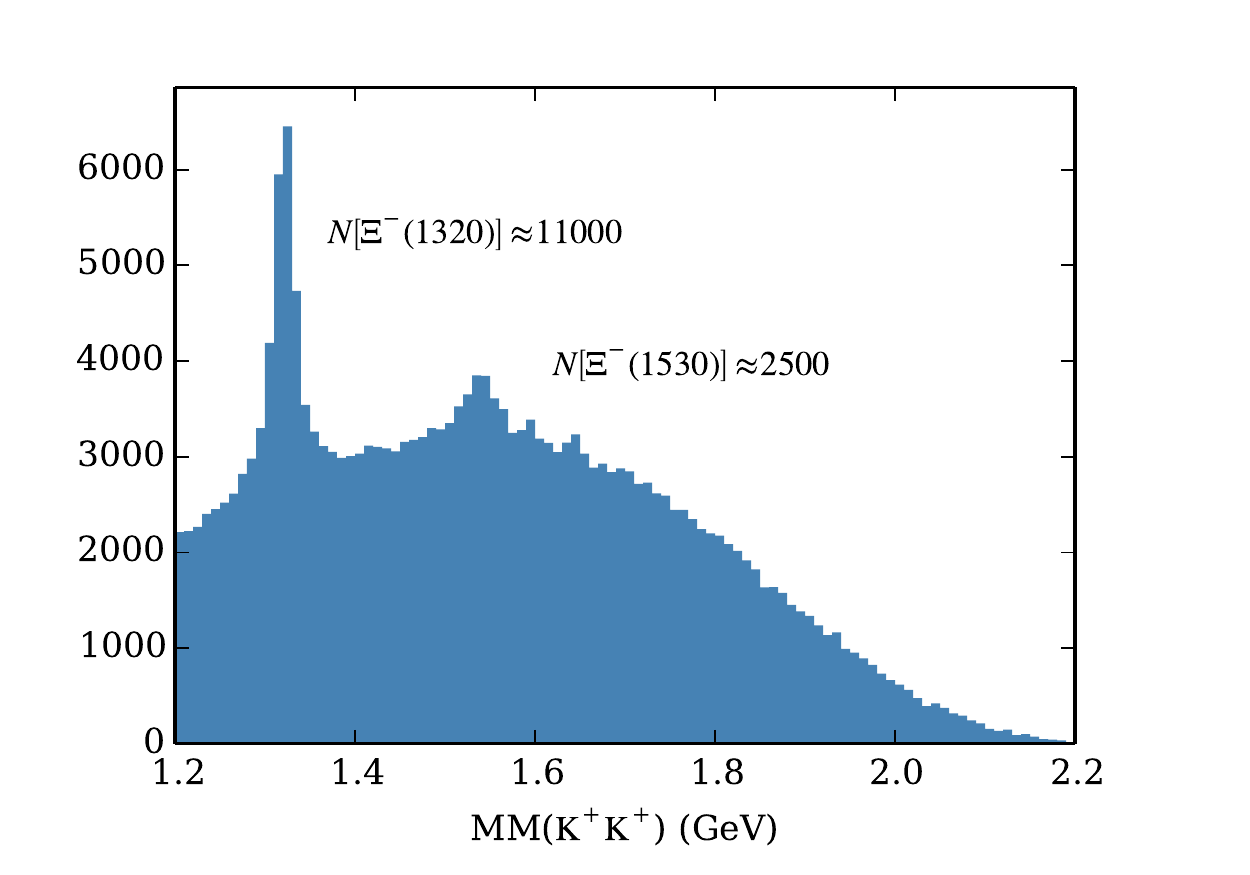}
\caption[MM(K$^+$K$^+$K$^-$)]{\label{fig:xim1320.dcs.kpkp.final.mmkk}
Missing mass off (K$^+$K$^+$) showing the $\Xi$ spectrum above a smooth 
background, summed over all angles and all $E_\gamma$. 
The two lowest-lying states are shown along with their approximate yields.
The missing-mass resolution of the CLAS detector is about 0.01 GeV.}
\end{center}\end{figure}

In the missing mass off $K^+ K^+$ 
(Fig.~\ref{fig:xim1320.dcs.kpkp.final.mmkk}), 
the strong peak at 1.32~GeV corresponds to the $\Xi$ ground state 
($J^P = \tfrac{1}{2}^{-}$) and the smaller peak at 1.53~GeV is the $\Xi^*$ 
first excited state ($J^P = \tfrac{3}{2}^{-}$). 
No other statistically significant structures are seen in this mass spectrum.

It may be somewhat surprising that no statisically significant peaks are 
seen corresponding to the known $\Xi^*$ states above the $\Xi$(1530). 
The most likely explanation is that the same reaction mechanism that leads 
to the $\Xi$ and $\Xi$(1530) do not, for photoproduction, extend to these 
higher-mass $\Xi$ states, which have different spin and parity. 
However, theoretical calculations and more precise measurements are needed 
to test this hypothesis.

\begin{figure}[htpb]\begin{center}
\includegraphics[width=1.0\columnwidth]{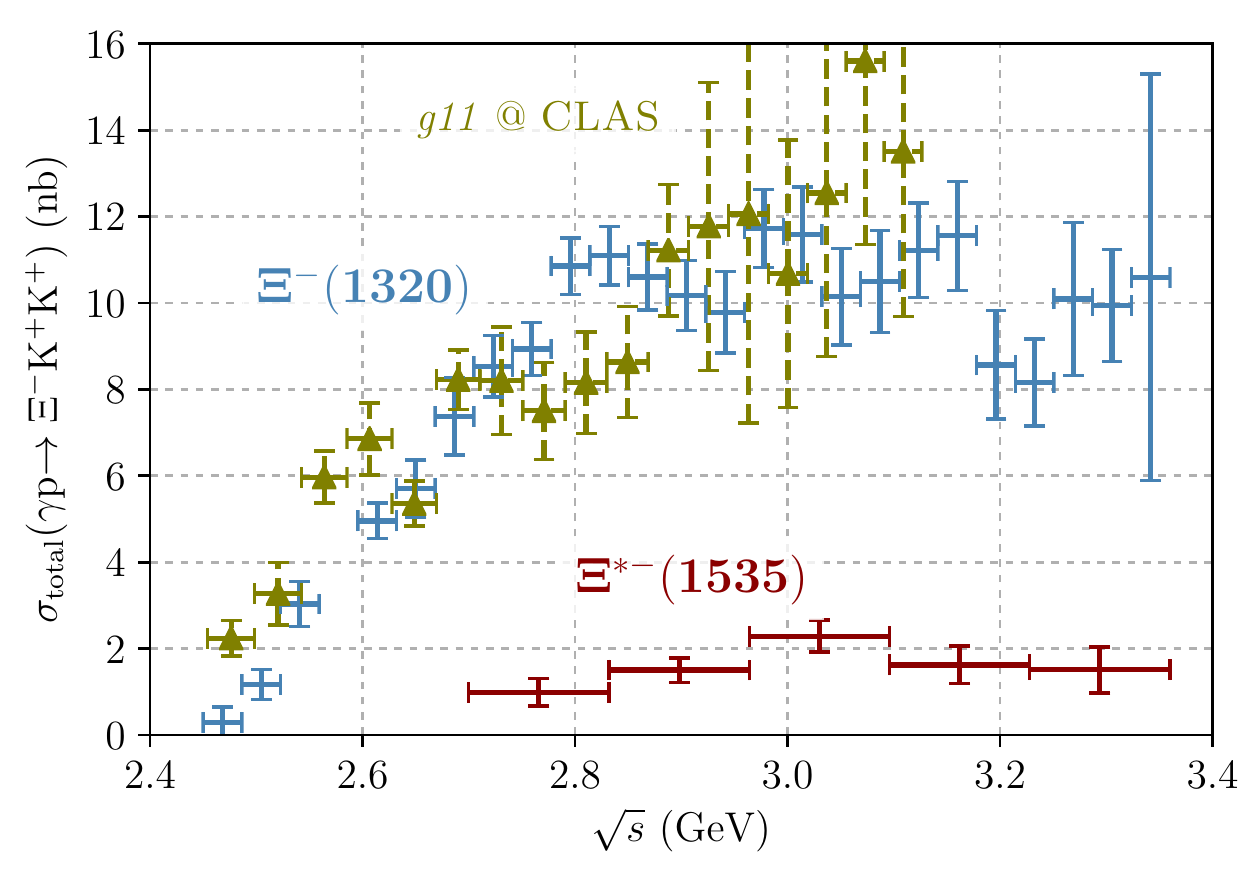}
\caption[Total Cross Section]{\label{fig:xim1320.tcs}
Total cross section of the $\Xi^-(1320)$ ground state and $\Xi^-(1535)$ 
first excited state, from photoproduction threshold to $W = 3.3$~GeV. 
The previous CLAS data (from the g11 dataset) with large uncertainties above 
2.9 GeV are shown by the triangle markers.}
    \end{center}
\end{figure}

The total cross sections are shown in Fig.~\ref{fig:xim1320.tcs}, together 
with previous CLAS results \cite{guo2007cascade} for the $\Xi(1320)$ ground 
state and the $\Xi(1535)$ first excited state.
This result was obtained by integrating fits to the angular distributions of 
differential cross sections, which will be shown in a forthcoming paper. 
These new data, for the first time, show that the total cross section levels 
off above $W \simeq 2.8$~GeV, which is only evident now with the new CLAS 
data. This suggests that the production mechanism is not from an intermediate 
$s$-channel resonance, in part because it is consistent with the predictions 
of a two-step reaction mechanism through intermediate $N^*$ and $Y^*$ 
resonances of Oh {\it et al.} \cite{nakayama2006photoproduction}, which 
becomes flat for $W > 2.9$ GeV.

If the total cross sections maintain reasonably constant values up to higher 
photon energies, then further studies of the $\Xi$ and $\Xi^*$ states could 
be done with the future CLAS12 detector \cite{stepanyan2010hadron}. 
Preliminary estimates \cite{clas12.ftag} show that more than a factor of ten 
times the statistics on $\Xi$ production could be obtained at CLAS12. 
In addition, the GlueX experiment at Jefferson Lab is expected to soon have 
similarly good statistics for $\Xi$ production at higher photon energies
\cite{ernst.2017.aps}.

The systematic uncertainties, which are not included in 
Fig.~\ref{fig:xim1320.tcs}, include 6\% due to the normalization 
(such as photon flux), 5\% due to integration of fits to the angular 
distributions, 3\% due to variations of cuts and detector acceptance, and 
3\% due to other effects such as target length and electronics livetime, 
giving 8.8\% overall.

\begin{figure}[htpb]\begin{center}
\includegraphics[width=1.0\columnwidth]{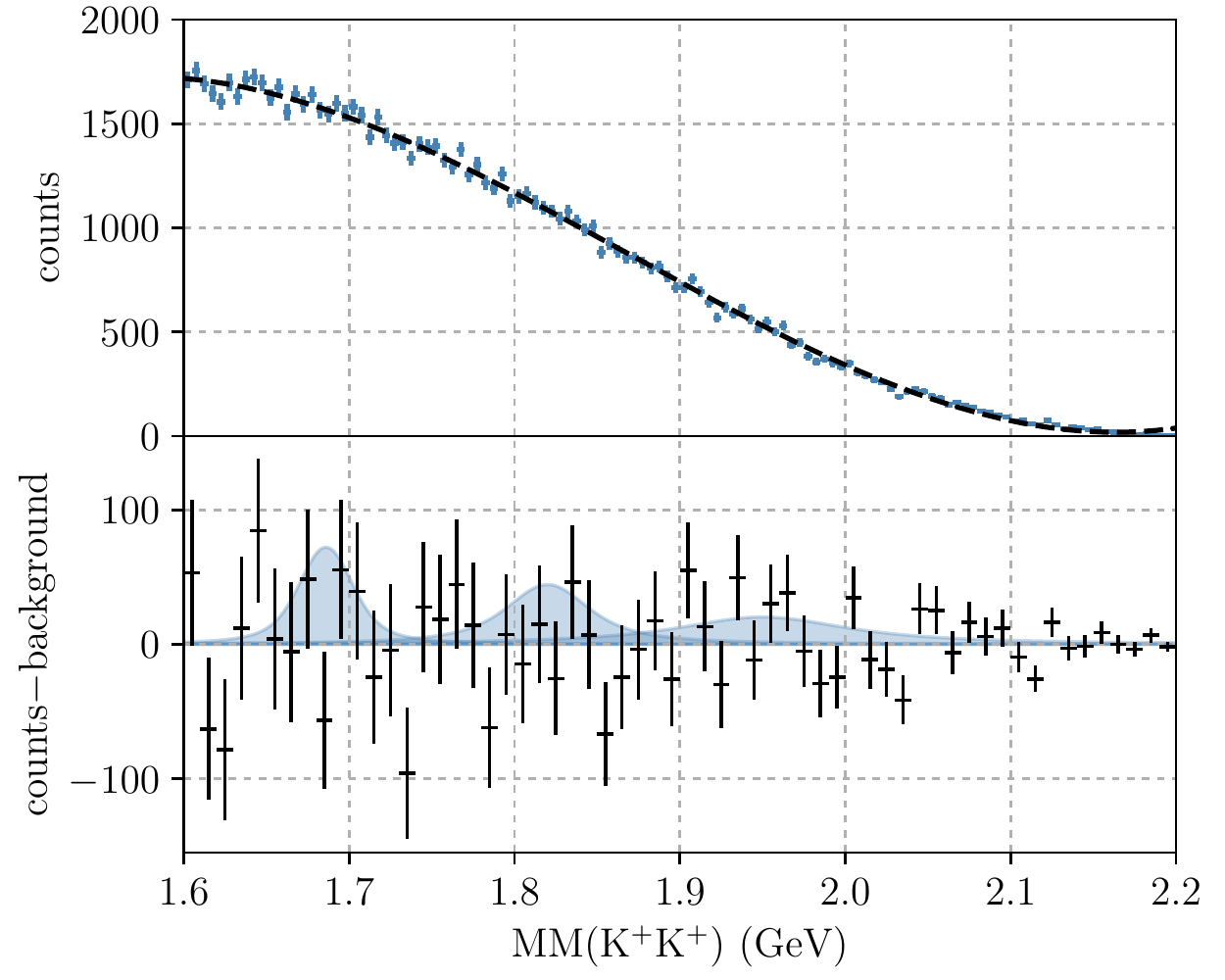}
\caption[]{\label{fig:ulimit.highxi}
Missing Mass off K$^+$K$^+$ with a fitted 3$^{rd}$-order polynomial background. 
The filled curves correspond to the 90\% confidence (Feldman-Cousins 
prescription) yield upper limits of $\Xi^{*-}$ states at 1690, 1820 and 
1950~MeV.}
\end{center}\end{figure}

No evidence is found for higher-mass $\Xi$ states in the missing mass off 
$K^+ K^+$ of this experiment shown in 
Fig.~\ref{fig:xim1320.dcs.kpkp.final.mmkk}. 
Upper limits were calculated on the production total cross sections of the 
three best-known excited states: the $\Xi(1690)$, the $\Xi(1820)$  and the 
$\Xi(1950)$ \cite{pdg} at 0.75 nb, 1.01 nb,  and 1.58 nb, respectively, at 
the 90\% confidence limit. 
Figure~\ref{fig:ulimit.highxi} shows an expansion of the missing-mass spectrum 
of Fig. \ref{fig:xim1320.dcs.kpkp.final.mmkk}. 
The spectrum is fit to a third-order polynomial along with three Voigtians with 
fixed means, Lorentzian-widths and Gaussian-widths for the $\Xi(1690)$, 
$\Xi(1820)$ and $\Xi(1950)$, using their measured widths \cite{pdg} 
shown by the filled curve for a 90\% confidence level upper limit. 

The ratio of the $\Xi(1690)$ to $\Xi(1530)$ cross sections in 
$\Sigma^-$ production was measured by WA89 \cite{adamovich1998first} 
to be approximately 2.2\%. 
Of course, the photoproduction mechanism of the CLAS experiment is different 
from WA89's largely hadronic process,
but the upper limit for the $\Xi(1690)$ from the CLAS data is consistent 
with this hadronic ratio.

In conclusion, we report the first total cross sections for 
photoproduction of the $\Xi(1320)$ and $\Xi(1530)$ ground states over 
center-of-mass energies above $W=2.8$~GeV, where the cross section 
is found to level off.  One of the 
goals of the present measurements was to explore the spectrum of 
excited $\Xi^*$ states, but surprisingly these states are much suppressed 
in photoproduction, and only upper limits could be determined for the 
total cross sections for three known $\Xi^*$ states 
(at masses 1690, 1820, and 1950 MeV).
The production mechanism that explains such small photoproduction cross 
sections is not yet known, and begs for an explanation from future theoretical 
calculations. 
More measurements at higher photon energies using the upgraded Jefferson 
Lab accelerator will soon be available to test such calculations.

As mentioned earlier, the spectrum of the $\Xi$ baryons is incomplete, since 
we expect one $\Xi^*$ resonance for each known $N^*$ resonance.  This paper 
shows that the photoproduction mechanism for $W < 3.3$ GeV does 
not strongly populate higher-mass $\Xi^*$ resonances, and so one must look 
to other methods to complete the spectrum of $\Xi^*$ resonances.

\begin{acknowledgments}
The authors gratefully acknowledge the work of Jefferson Lab staff in the 
Accelerator and Physics Divisions. 
This work was supported by: 
the United Kingdom’s Science and Technology Facilities Council (STFC); 
the Chilean Comisi\`on Nacional de Investigaci\`on Cient\`ifica y Tecnoĺ\`ogica (CONICYT); 
the Italian Istituto Nazionale di Fisica Nucleare; the French Centre National de la Recherche
Scientifique; the French Commissariat \`a l’Energie Atomique;
the U.S. National Science Foundation; and the National
Research Foundation of Korea. Jefferson Science Associates,
LLC, operates the Thomas Jefferson National Accelerator
Facility for the the U.S. Department of Energy under Contract
No. DE-AC05-06OR23177. 
\end{acknowledgments}

\clearpage

\end{document}